\UseRawInputEncoding 
\documentclass[twocolumn,preprintnumbers,amsmath,superscriptaddress,amssymb,aps] {revtex4}
\usepackage{graphicx}
\usepackage{dcolumn}
\usepackage{bm}

\begin{document}
\title{High temperature ideal Weyl semimetal phase and Chern insulator phase in
%double perovskite
ferromagnetic BaEuNiOsO$_{6}$ and its (111) (BaEuNiOsO$_6$)/(BaTiO$_3$)$_{10}$ superlattice}
\par
\author{Hai-Shuang Lu}
\address{School of Electronic and Information Engineering, Changshu Institute of Technology, Changshu 215500, People's Republic of China}
\author{Guang-Yu Guo}
\email{gyguo@phys.ntu.edu.tw}
\address{Department of Physics, National Taiwan University, Taipei 10617, Taiwan}
\address{Physics Division, National Center for Theoretical Sciences, Taipei 10617, Taiwan}
%\date{\today}

\begin{abstract}
Weyl semimetals (WSMs) have recently stimulated intensive interest because
they exhibit fascinating physical properties and also promise exciting
technological applications. So far, however, the few confirmed magnetic WSMs
generally have a large number of Weyl points either located away from the Fermi level
($E_F$) or shrouded by nontopological Fermi surface pockets.
Based on first-principles density functional theory calculations,
we establish cubic double perovskite BaEuNiOsO$_{6}$ to be a high Curie temperature ($T_c$)
ferromagnetic WSM with magnetization along the [111] direction,
just two pairs of Weyl points at the $E_F$ and $T_c = 325$ K.
The strong ferromagnetism is attributed to the strong ferromagnetic Ni 3$d$-Eu 4$f$-Os 5$d$
coupling induced by the substitution of half of Ba atoms with Eu atoms in
double perovskite Ba$_2$NiOsO$_{6}$.
Moreover, the momentum separation of one Weyl point pair is large,
thus giving rise to not only a long (001) surface Fermi arc but also
large anomalous Hall conductivity.
Intriguingly, as a unique physical result of a ferromagnetic WSM, the (111) BaEuNiOsO$_{6}$
monolayer superlattice (BaEuNiOsO$_{6}$)/(BaTiO$_3$)$_{10}$, being its (111) quantum-well structure,
is found to be a high temperature ($T_c = 210$ K) Chern insulator with a large band gap of $\sim$90 meV.
%as expected because it is the (111) quantum-well structure of bulk BaEuNiOsO$_{6}$.
Therefore, cubic double perovskite BaEuNiOsO$_{6}$ will provide
a superior high temperature material plotform for exploring fundamental physics of Weyl fermions
and its (111) monolayer superlattices will offer a high temperature magnetic topological
insulator for studying exotic quantum phenomena such as quantum anomalous Hall effect.
\end{abstract}

\maketitle

\section{Introduction}
Weyl semimetals (WSMs), hosting Weyl fermions described by the
Weyl Hamiltonian~\cite{Weyl29}, have attracted extensive interest
due to their unusual physical properties and potential
technological applications~\cite{Wanxg2011, Shiw2018,
Volovik,YanB2017,ArmitageNP2018,Hasanmz2010}. In a WSM, the
conduction and valence bands disperse linearly through the Weyl
points in the Brillouin zone (BZ). The Weyl points behave as
the monopoles of Berry curvature with opposite chiralities and
connect the two ends of the nonclosed surface Fermi arcs~\cite{DingH2015,
YangLX2015, LiuZK2016, Hasanmz2015, Shim2016, Inoueh2016}.
Magnetic WSMs with broken time-reversal symmetry are especially
interesting. They provide a playground for the interplay among
magnetism, symmetry and topology, which can
give rise to exotic quantum phenomena such as chiral
magnetic effect~\cite{HuangChen2015,ZhangJia2016},
large universal anomalous Hall effect (AHE)~\cite{LiuFelser2018}, low-frequency divergent
bulk photovoltaic effect~\cite{Ahn2020} and also
quantum anomalous Hall effect (QAHE)~\cite{Xug2011}. Although a
number of magnetic WSMs candidates have been proposed, e.g,
pyrochlore iridate Y$_{2}$Ir$_{2}$O$_{7}$~\cite{Wanxg2011}, spinel
compounds HgCr$_{2}$Se$_{4}$~\cite{Xug2011} and
VMg$_{2}$O$_{4}$~\cite{TonyLow}, half-Heusler $R$PtBi ($R$ = Gd and
Nd)~\cite{Hirschberger2016,Shekhar2018}, certain Co$_{2}$-based
Heusler compounds~\cite{KublerJ2016, Wangz2016,
Changg2016,HasanMZ19}, ferromagnetic (FM) kagome-lattice
Co$_{3}$Sn$_{2}$S$_{2}$~\cite{LiuFelser2018,ChenYL19,Beidenkopf19},
and noncentrosymmetric $R$AlGe family of compounds ($R$ = rare
earth elements)~\cite{ChangG2018,Sanchez2020}, few have been confirmed by
experiments~\cite{HasanMZ19,ChenYL19,Beidenkopf19,Sanchez2020}. Furthermore,
in the few confirmed WSMs, there are usually a large number of Weyl points
located either far above or below the Fermi level and also shrouded by
nontopological Fermi surface pockets~\cite{LiuFelser2018,ChangG2018,HasanMZ19}.
This certainly hinders the studies of the exotic properties especially
novel transport phenomena in WSMs. Therefore, it is highly
desirable to find high Curie temperature ($T_c$) magnetic WSMs with
only a couple of pairs of Weyl points located at or close to the Fermi level
(so-called ideal WSMs)~\cite{Soh2019}.

Transition metal oxide perovskites (ABO$_{3}$) and also double
perovskites A$_{2}$BB$'$O$_{6}$ possess high cubic
symmetries and exhibit a rich variety of fascinating properties
such as colossal magnetoresistance and half-metallic behavior~\cite{Kobayashi}.
Their atomic scale heterostructures offer the prospect of further enhancing these
fascinating properties or of combining them to realize exotic
properties and functionalities~\cite{Mannhart,Hwang12}.
Recently, based on their tight-binding (TB) modelling and density functional theory (DFT) calculations,
Xiao {\it et al.} showed in their seminal paper~\cite{Xiao11} that various quantum topological phases
could appear in a class of (111) perovskite bilayers (BLs) sandwiched by insulating
perovskites where transition metal atoms in the BLs form a buckled honeycomb lattice.
Subsequently, the electronic structures of a large number of (111) perovskite BLs
and also double-perovskite monolayers (MLs) in the (111) oxide superlattices
were investigated and some of them were predicted to host
quantum spin Hall, quantum anomalous Hall (QAH) and other topological phases
(see, e.g., Ref. ~\cite{Weng15} and references therein).
In particular, (111) perovskite BLs~\cite{Ran11,Dong15,Dai15,Held17,Chandra17,Lu19}
and also (111) double perovskite MLs~\cite{Cook14,Baidya16}
have been predicted to be QAH insulators.

However, very few magnetic WSMs based on oxide perovskite
or double perovskite structures have been reported~\cite{ChenY2013},
despite a lot of magnetic WSMs based other materials have been predicted, as mentioned above.
We notice that a magnetic WSM can be viewed as a stack of
two-dimensional QAH insulators with strong coupling in the
stacking direction~\cite{Wanxg2011,Xug2011,Yang2011,Weng15}.
This naturally leads to a "inverse engineering" strategy
for finding magnetic WSM phases in bulk perovskites and doube perovskites.
One first looks for high $T_c$ QAH phases in
perovskite BLs or double perovskite MLs.
One then examines possible  magnetic WSM phases in the corresponding
bulk perovskites and doube perovskites.

In this work, as a continuation of our endeavor in
searching for high $T_c$ QAH phases in perovskite BLs~\cite{Lu19}
and also in double perovskite MLs~\cite{LuGuo19},
we study the electronic structure of the (111) ML of Eu-doped double perovskite Ba$_2$NiOsO$_6$.
Double perovskite Ba$_2$NiOsO$_6$ is a rare FM semiconductor with $T_{c}$ of
100 K~\cite{Feng16}. Our previous DFT calculation revealed that
the ferromagnetism is driven by the FM coupling between Ni and
Os atoms~\cite{LuGuo19}, which is very rare between the B and
B$'$ in double perovskites~\cite{Wang09}.
Furthermore, our previous calculation showed that
its energy bands near the Fermi level are dominated by
Os 5$d$ $t_{2g}$ orbitals with $t_{2g}^{2}$ electron
configuration~\cite{LuGuo19}. Consequently, the strong spin-orbit coupling
(SOC) on the heavy Os atom opens the semiconducting gap,
thus making Ba$_2$NiOsO$_6$ a so-called Dirac semiconductor~\cite{Feng16,LuGuo19}
According to Ref. ~\cite{Xiao11},
the (111) double perovskite MLs with
$t_{2g}^{1}$, $t_{2g}^{2}$, $t_{2g}^{3}$ and $t_{2g}^{4}$
electronic configurations are all possible topological materials.
Previously, we found that the first band gap below the valence band maximum
in the (111) Ba$_2$NiOsO$_6$ ML is topologically nontrivial,
although it is an ordinary FM semiconductor
with a predicted $T_c$ of 70 K~\cite{LuGuo19}.
Now, if one A-site atom in Ba$_2$NiOsO$_6$ is replaced by a magnetic 4$f$ atom,
forming additional FM coupling channels, its ferromagnetism could be
further strengthened, resulting in a higher $T_c$.
Moreover, such A-site substitution would break
the spatial inversion symmetry of both bulk Ba$_2$NiOsO$_6$ and its (111) ML,
thus leading to the possible emergence of topological phases.

\begin{figure}
\includegraphics[width=8cm]{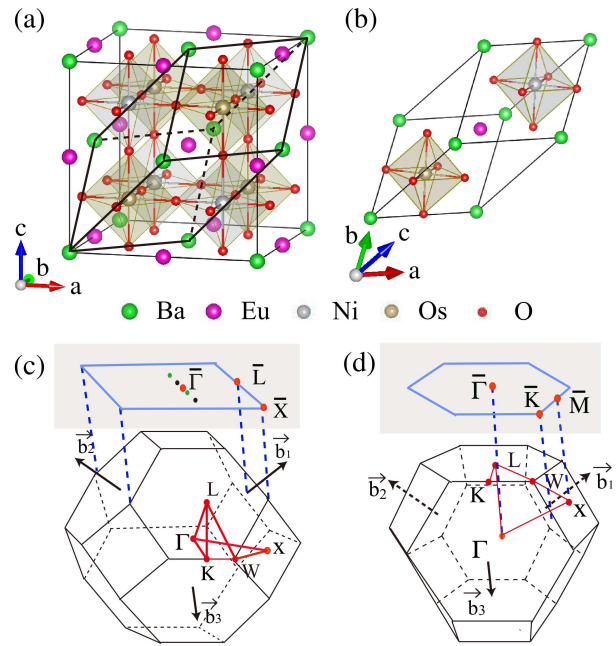}
\caption{
Cubic crystal structure of bulk BaEuNiOsO$_6$. (a) Conventional
cubic cell and (b) fcc primitive cell as well as the associated
fcc Brillouin zone (BZ) with (c) the corresponding (001) surface BZ
and (d) the corresponding (111) surface BZ.
In (a), the black lines indicate the fcc primitive cell (b).
In (c) and (d),
$\vec{b}_{1}$, $\vec{b}_{2}$ and $\vec{b}_{3}$ are
the reciprocal lattice vectors. }
\end{figure}

\begin{figure}
\includegraphics[width=8cm]{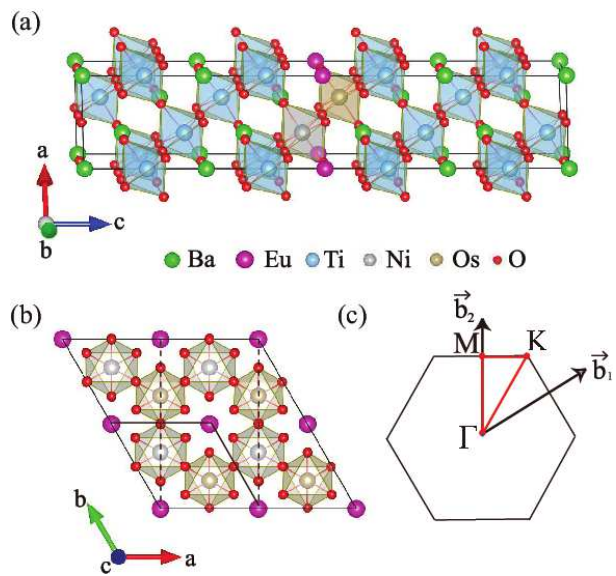}
\caption{Crystal structure of the (111) (BaEuNiOsO$_6$)/(BaTiO$_3$)$_{10}$
superlattice. (a) Side view along the $b$-axis, (b) top view along the $c$-axis
and (c) the corresponding two-dimensional (2D) BZ.
In (b), the black lines denote the 2D primitive cell.
In (c), $\vec{b}_{1}$ and $\vec{b}_{2}$ represent the 2D reciprocal lattice vectors.
}
\end{figure}

Indeed, as will be presented in Sec. III, our DFT calculations indicate
that the (111) (BaEuNiOsO$_{6}$)/(BaTiO$_{3}$)$_{10}$
ML superlattice is a Chern insulator with high $T_{C}$ of 210 K.
Furthermore, our calculations also reveal that bulk BaEuNiOsO$_{6}$
is a FM WSM with high $T_{C}$ of 325 K and two pairs of Weyl points at the Fermi level only.
In other words, guided by the inverse engineering strategy, we discover that bulk BaEuNiOsO$_{6}$
is a rare ideal ferromagnetic WSM. Therefore, double-perovskite BaEuNiOsO$_{6}$
will provide a superior platform for exploring fascinating Weyl fermion physics
and its (111) ML superlattice and its (111) ML superlattice will be
a high temperature Chern insulator for studying exotic quantum phenomena such as QAH effect.
The rest of this paper is organized as follows.
In the next section, a brief description of the crystalline structures of
bulk BaEuNiOsO$_{6}$ and its (111) ML superlattice as well as the used theoretical methods
and computational details is given. In Sec. III, the calculated magnetic properties and
electronic band structures as well as the uncovered quantum topological phases are presented.
Finally, the conclusions drawn from this work are summarized in section IV.

\begin{figure}
\includegraphics[width=7cm]{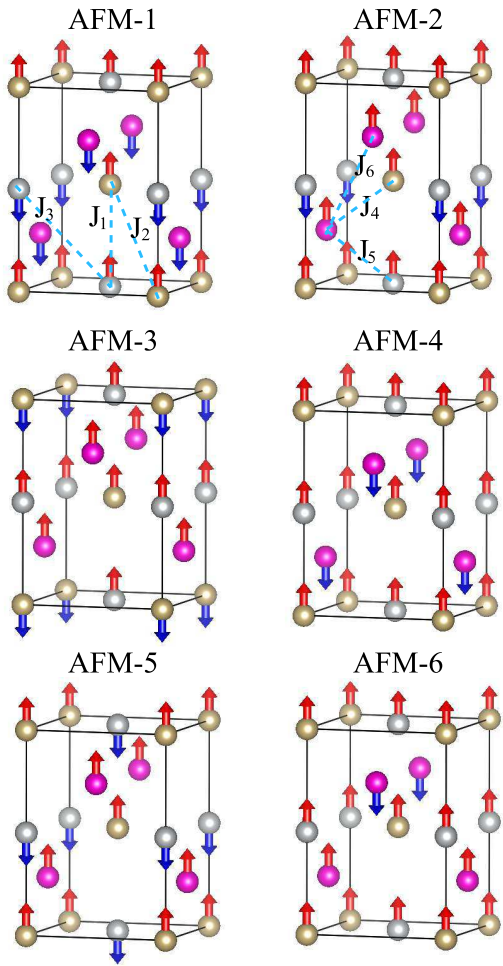}
\caption{Six considered antiferromagnetic configurations (AFM-${i}$) in bulk
BaEuNiOsO$_6$.}
\end{figure}

\begin{figure}
\includegraphics[width=6.5cm]{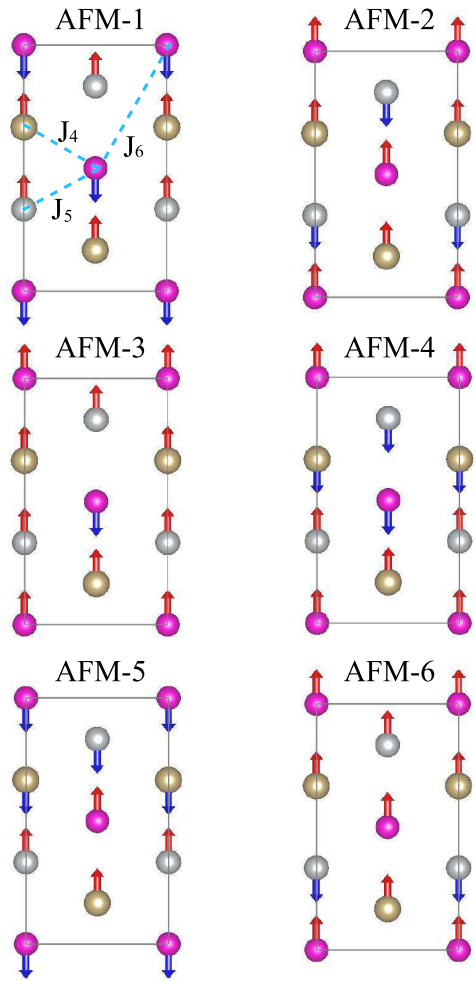}
\caption{Six considered antiferromagnetic configurations (AFM-${i}$)
in the (111) BaEuNiOsO$_{6}$ monolayer in the (111) (BaEuNiOsO$_{6}$)/(BaTiO$_{3}$)$_{10}$
superlattice. The cell here is indicated by the dashed lines in Fig. 2(b).}
\end{figure}

\section{THEORY AND COMPUTATIONAL DETAILS}

%\begin{table}%[h]
%\caption{Wyckoff-positions of cubic BaEuNiOsO$_{6}$ with space group F$\bar{4}$3m (No. 216) and its (111) monolayer with
%space group $P3m1$ ($C_{3v}^1$) (no. 156). The Lattice parameter is 8.006 (\AA) for cubic BaEuNiOsO$_{6}$, and that for (111) monolayer
%is a=b= 5.6962 (\AA) and c = 27.66 (\AA).}
%\begin{ruledtabular}
%\begin{tabular}{cccccc}
%& & Cubic BaEuNiOsO$_{6}$ & & \\
%\hline
% Atom & Site  & x     &  y     & z \\
%\hline
% Ba    & 4c    &0.250000   &0.250000    &0.250000      \\
% Eu    & 4d    &0.750000   &0.750000    &0.750000      \\
% Os    & 4b    &0.500000    &0.500000    &0.500000      \\
% Ni    & 4a    &0.000000    &0.000000    &0.000000      \\
% O     & 24f   &0.256742    &0.000000    &0.000000   \\
%\hline
% & & The (111) monolayer & & \\
%\hline
% Atom & Site  & x     &  y     & z \\
%\hline
% Ba    & 3d    &0.333333   &0.666667    &0.582496  \\
% Eu    & 1a    &0.000000   &0.000000    &0.497787  \\
% Os    & 1c    &0.666667   &0.333333    &0.540598  \\
% Ni    & 1b    &0.333333   &0.666667    &0.457222  \\
% O     & 6e    &0.332910   &0.166455    &0.580242  \\
% &        & (0.989495) & (0.494747) & (0.501347) \\
%\end{tabular}
%\end{ruledtabular}
%\end{table}

We consider face-centered-cubic double perovskite BaEuNiOsO$_{6}$ with
noncentrosymmetric space group F$\bar{4}$3m (No. 216) (see Fig. 1),
and also its (111) ML sandwiched by
an insulating perovskite BaTiO$_3$ slab as in the
(BaEuNiOsO$_6$)/(BaTiO$_3$)$_{10}$ superlattice grown along
the [111] direction (see Fig. 2). The resultant ML
superlattice is a trigonal structure [space group $P3m1$ ($C_{3v}^1$), no. 156]
without spatial inversion symmetry.
In each (111) BaEuNiOsO$_{6}$ ML, Ni and Os atoms
form a buckled honeycomb lattice separated by an EuO$_{3}$ layer.
The Brillouin zones of bulk BaEuNiOsO$_{6}$ and the (111)
ML are shown in Figs. 1(c) and 1(d) and Fig. 2(c), respectively.
Clearly, the latter is the folded BZ of the former along the $\Gamma$-L direction.
Since the BaTiO$_{3}$ slab in the
(BaEuNiOsO$_{6}$)/(BaTiO$_{3}$)$_{10}$ superlattice is much
thicker than the BaEuNiOsO$_6$ ML, the BaTiO$_{3}$ slab
could be regarded as the substrate. Therefore, in the present
structural optimization calculation, the in-plane
lattice constant is fixed at ${\sqrt{2}}$${a_{0}}$ = 5.6962 \AA,
where $a_{0}=4.0278$ \AA$ $ is the theoretically determined lattice constant
of cubic BaTiO$_3$, while the lattice constant $c$ and
the internal coordinates of all the atoms in the superlattice are
then optimized theoretically.
%The Wyckoff-positions for bulk BaEuNiOsO$_{6}$ and its (111) ML are given in Table I.
Also, the theoretically determined lattice parameters and atom
positions for both cases are shown, respectively, in Tables S1 and S2
for bulk BaEuNiOsO$_{6}$ and also in Table S3 for the
(BaEuNiOsO$_{6}$)/(BaTiO$_{3}$)$_{10}$ superlattice in the
Supplementary Materials (SM)~\cite{SM}. It is worth noting
that the primitive unit cell [see Fig. 1(b)] is used in all the
present calculations for bulk cubic BaEuNiOsO$_{6}$.

The present electronic and magnetic structure
as well as structural optimization calculations are based on the
DFT with the generalized gradient
approximation (GGA)~\cite{Perdew96}. The accurate
projector-augmented wave (PAW) method~\cite{PEB}, as implemented
in the Vienna {\it ab initio} simulation package
(VASP)~\cite{Kresse93}, is used. The relativistic PAW potentials
are adopted in order to include the SOC. The valence
configurations of Ba, Eu, Ni, Os, Ti and O atoms adopted in the
present calculations are
5\emph{s}$^{2}$5\emph{p}$^{6}$6\emph{s}$^{2}$,
5\emph{s}$^{2}$5\emph{p}$^{6}$4\emph{f}$^{7}$6\emph{s}$^{2}$,
3\emph{p}$^{6}$3\emph{d}$^{8}$4\emph{s}$^{2}$,
5\emph{p}$^{6}$5\emph{d}$^{6}$6\emph{s}$^{2}$,
3\emph{s}$^{2}$3\emph{p}$^{6}$3\emph{d}$^{2}$4\emph{s}$^{2}$ and
2\emph{s}$^{2}$2\emph{p}$^{4}$, respectively. To better account
for the on-site electron correlation on the Os 5\emph{d}, Ni
3\emph{d} and Eu 4\emph{f} shells, the GGA+U
method~\cite{dudarev98} is adopted with the effective Coulomb
repulsion energies $U_{Os}$ = 2.0 eV, $U_{Ni}$ = 5.0 eV, and
$U_{Eu}$ = 4.0 eV. These $U$ values were found appropriate for
perovskite oxide EuTiO$_3$~\cite{lu15} and double-perovskite oxide
Ba$_{2}$NiOsO$_6$~\cite{LuGuo19} in our earlier works~\cite{LuGuo19, lu15}.
In particular, the GGA+U calculation with $U_{Eu}$ = 4.0 eV correctly
predicted bulk EuTiO$_3$ to be a G-type antiferromagnetic insulator
with the Neel temperature $T_N = 6.0$ K.~\cite{lu15} A large plane-wave cutoff of 400 eV
and the small total energy convergence criterion of 10$^{-5}$ eV
are used throughout. Fine Monkhorst-Pack \emph{k}-meshes of
50$\times$50$\times$50 and 10$\times$10$\times$2 are used for
the BZ integrations for bulk BaEuNiOsO$_6$ and the ML superlattice, respectively.
In the structural optimizations, ferromagnetic GGA+U scalar-relativistic calculations
are performed.

To find the ground state magnetic configuration and to understand
the magnetic interactions in bulk Ba$_{2}$NiOsO$_6$ and its (111) ML,
we consider seven possible magnetic structures for each system including the
ferromagnetic structure and six antiferromagnetic (AFM) structures as
shown in Figs. 3 and 4, respectively. One can then evaluate the
nearest-neighbor Ni-Os (\emph{J$_{1}$}), Os-Os (\emph{J$_{2}$}),
Ni-Ni (\emph{J$_{3}$}), Eu-Os (\emph{J$_{4}$}), Eu-Ni
(\emph{J$_{5}$}) and Eu-Eu (\emph{J$_{6}$}) magnetic coupling
parameters by mapping the calculated total energies of the seven
magnetic configurations to the classical Heisenberg model
$H = E_{0} - \sum_{i>j} J_{ij}(\hat{e}_{i}\cdot\hat{e}_{j})$,
where $J_{ij}$ is the exchange coupling parameter between
sites \emph{i} and \emph{j}, and $\hat{e}_{i}$ is the unit vector indicating the
direction of spin on site \emph{i} (see supplementary note 1 in the SM~\cite{SM}).
Based on the calculated $J$
values, we could estimate magnetic ordering temperature ($T_c$)
within a mean-field approximation given by $ k_BT_c=\frac {1}{3}
\sum_i z_i J_i$ where $z_i$ are the numbers of Ni-Os, Os-Os, Ni-Ni, Eu-Os, Eu-Ni
and Eu-Eu pairs (bonds) in the considered system (see supplementary note 1 in the SM).

\begin{table}\footnotesize
\caption{\label{tab:table1} Calculated total energies ($E_{FM}$, $E_{AFM-i}$)
of the FM and six considered AFM states (see Figs. 3 and 4) with the GGA+U
method,
exchange coupling parameters ($J_i$) and also the estimated magnetic ordering
temperature ($T_{c}$) of bulk BaEuNiOsO$_6$ and its (111) monolayer in
the (111) (BaEuNiOsO$_{6}$)/(BaTiO$_{3}$)$_{10}$ superlattice.
Here $J_{1}$ ($d_{Ni-Os}$), $J_{2}$ ($d_{Os-Os}$), $J_{3}$ ($d_{Ni-Ni}$),
$J_{4}$ ($d_{Eu-Os}$), $J_{5}$ ($d_{Eu-Ni}$), and $J_{6}$ ($d_{Eu-Eu}$)
represent the nearest Ni-Os, Os-Os, Ni-Ni, Eu-Os, Eu-Ni and Eu-Eu
exchange coupling parameter (interatomic distance), respectively.}
\begin{ruledtabular}
\begin{tabular}{cccc}
                  & Bulk    &   (111) monolayer    \\
\hline
$E_{FM}$    (meV/f.u.)    &   0       &   0      \\
$E_{AFM-1}$ (meV/f.u.)    &   133.325   & 28.39   \\
$E_{AFM-2}$ (meV/f.u.)    &   51.55     & 31.99    \\
$E_{AFM-3}$ (meV/f.u.)    &   106.075   & 4.575     \\
$E_{AFM-4}$ (meV/f.u.)    &   105.26    & 38.84    \\
$E_{AFM-5}$ (meV/f.u.)    &   105.695   & 42.90   \\
$E_{AFM-6}$ (meV/f.u.)    &   26.085   & 18.455    \\
$d_{Ni-Os}$ (\AA)       &   3.965     & 4.016    \\
$d_{Os-Os}$ (\AA)       &   5.607     & 5.696   \\
$d_{Ni-Ni}$ (\AA)       &   5.607     & 5.696   \\
$d_{Eu-Os}$ (\AA)       &   3.434     & 3.495   \\
$d_{Eu-Ni}$ (\AA)       &   3.434     & 3.475   \\
$d_{Eu-Eu}$ (\AA)       &   5.607     & 5.696    \\
$J_{1}$ (meV)           &   8.186     & 3.981    \\
$J_{2}$ (meV)           &   1.492     & 6.468    \\
$J_{3}$ (meV)           &  -0.162     & 0.615   \\
$J_{4}$ (meV)           &   10.222     & 3.381    \\
$J_{5}$ (meV)           &  -0.102     & 1.351    \\
$J_{6}$ (meV)           &  -3.121    & -2.405  \\
$T_{c}$   (K)           &$\sim$325 ($\sim$100\footnotemark[1])    &$\sim$210 ($\sim$69\footnotemark[2])\\
\end{tabular}
\end{ruledtabular}
\footnotemark[1]{Experimental value of bulk Ba$_{2}$NiOsO$_6$ from
Ref. [~\onlinecite{Feng16}].}\\
\footnotemark[2]{Calculated value for (111) Ba$_{2}$NiOsO$_6$
monolayer from Ref. [~\onlinecite{LuGuo19}].}
\end{table}

\begin{table}%[h]
\caption{Calculated spin ($m_s$) and orbital ($m_o$) magnetic moments as
well as band gap ($E_g$) and anomalous Hall conductivity ($\sigma_{xy}^A$)
of FM bulk BaEuNiOsO$_6$ with magnetization along the (111) direction and
its (111) monolayer in the (111) (BaEuNiOsO$_{6}$)/(BaTiO$_{3}$)$_{10}$ superlattice
with the perpendicular magnetization using the GGA+U+SOC method.
Also listed are the Eu $f$- as well as Os and Ni $d$-shell configurations.
For bulk BaEuNiOsO$_6$, $\sigma_{xy}^A$ is calculated at the energy level of the Weyl points.}
\begin{ruledtabular}
\begin{tabular}{cccc}
                                & Bulk  &(111) monolayer \\
  \hline
Eu$_{vc}$                     & 4$f^{7}$                   &4$f^{7}$             \\
Os$_{vc}$                     & 5$d^{2}$ ($t_{2g}^{2}$)    &5$d^{2}$ ($t_{2g}^{2}$)      \\
Ni$_{vc}$                     & 3$d^{8}$ ($t_{2g}^{6}$$e_{g}^{2}$) &3$d^{8}$ ($t_{2g}^{6}$$e_{g}^{2}$) \\
\emph{m$_{s}^{Os}$} ($\mu_B$/atom) & 1.499 &1.463  \\
\emph{m$_{o}^{Os}$} ($\mu_B$/atom) &-0.367 &-0.345 \\
\emph{m$_{s}^{Ni}$} ($\mu_B$/atom) & 1.768 &1.747  \\
\emph{m$_{o}^{Ni}$} ($\mu_B$/atom) & 0.213 & 0.209 \\
\emph{m$_{s}^{Ti}$} ($\mu_B$/atom) &- & 0.012  \\
\emph{m$_{o}^{Ti}$} ($\mu_B$/atom) &- & -0.003 \\
\emph{m$_{s}^{O}$} ($\mu_B$/atom)  &0.095  & 0.100  \\
\emph{m$_{o}^{O}$} ($\mu_B$/atom)  &-0.061 & -0.014 \\
\emph{m$_{s}^{Eu}$} ($\mu_B$/atom) &6.717  &6.670  \\
\emph{m$_{o}^{Eu}$} ($\mu_B$/atom) &-0.473 &-0.566  \\
\emph{m$_{t}^{Eu}$} ($\mu_B$/atom) & 6.244 &6.104  \\
\emph{m$_{t}^{Os}$} ($\mu_B$/atom) & 1.132 &1.118  \\
\emph{m$_{t}^{Ni}$} ($\mu_B$/atom) & 1.981 &1.956  \\
$E_g$ (eV)                         &0     &0.093  \\
$\sigma_{xy}^A$ ($e^{2}/hc$) & -1.122 & -1.000 \\
\end{tabular}
\end{ruledtabular}
\end{table}

The anomalous Hall conductivity (AHC) is calculated based on the
Berry-phase formalism~\cite{XiaoD10}. Within this Berry-phase
formalism, the AHC ($\sigma_{ij}^{A} = J^c_i/E_j$) is given as a
BZ integration of the Berry curvature for all the occupied
(valence) bands,
\begin{eqnarray}
\sigma_{xy}^{A} = -\frac{e^2}{\hbar}\sum_{n \in VB} \int_{BZ}\frac{d{\bf k}}{(2\pi)^3}\Omega_{xy}^n({\bf k}),\nonumber \\
\Omega_{xy}^n({\bf k}) = \nabla_{\bf k}\times\langle u_{n{\bf k}} | i\nabla_{\bf k} | u_{n{\bf k}} \rangle,
\end{eqnarray}
where ${\Omega_{ij}^n({\bf k})}$ and $u_{n{\bf k}}$ are, respectively,
the Berry curvature and cell periodic wave function for the $n$th band at ${\bf k}$.
$J^c_i$ is the $i$-component of the
charge current density ${\bf J}^c$ and $E_j$ is the $j$-component
of the electric field ${\bf E}$. Since a large number of $k$
points are needed to get accurate AHCs, we use the efficient
Wannier interpolation method~\cite{WangX06, LopezMG} based on
maximally localized Wannier functions (MLWFs)~\cite{MarzariN}.
Since the energy bands around the Fermi level are dominated by Os
5d $t_{2g}$ and Eu 4$f$ orbitals, 24 MLWFs are constructed by
fitting to the GGA+U+SOC band structures.
The band structures obtained by the Wannier interpolation agree
well with that from the GGA+U+SOC calculations (see Fig. S1 in the SM).
The AHC ($\sigma_{xy}^A$) is
then evaluated by taking very dense $k$-point meshes of 250$\times$
250$\times$250 and 200$\times$200$\times$2 for
bulk BaEuNiOsO$_{6}$ and its (111) ML, respectively. The surface states of bulk  Ba$_{2}$NiOsO$_6$
and the edge states of the (111) Ba$_{2}$NiOsO$_6$ ML are
calculated by the Green function technique~\cite{Rubio84, Rubio85}, as implemented
in the WannierTools code~\cite{Soluyanov18}.

\section{Results and discussion}
\subsection{Magnetic properties}
The calculated total energies of the seven considered magnetic configurations for both
systems are listed in Table I. These energies are calculated by the GGA + U method
with $U_{Os}$ = 2.0 eV, $U_{Ni}$ = 5.0 eV, and $U_{Eu}$ = 4.0 eV.
For both structures, the FM configuration
is the ground state. As mentioned before,
using these total energies, we evaluate the exchange
coupling parameters \emph{J$_{1}$}, \emph{J$_{2}$},
\emph{J$_{3}$}, \emph{J$_{4}$}, \emph{J$_{5}$} and \emph{J$_{6}$} (see supplementary note 1 in the SM~\cite{SM}),
as listed in Table I together with the corresponding atom-atom distances. For bulk BaEuNiOsO$_6$, the obtained
\emph{J$_{1}$}, \emph{J$_{2}$} and \emph{J$_{4}$} are positive (ferromagnetic), while
\emph{J$_{3}$}, \emph{J$_{5}$} and \emph{J$_{6}$} are negative (antiferromagnetic).
Furthermore, the magnitudes of \emph{J$_{1}$} and \emph{J$_{4}$}
are much larger than that of all other parameters including \emph{J$_{3}$}, \emph{J$_{5}$}
and \emph{J$_{6}$}. This means that ferromagnetism in bulk BaEuNiOsO$_6$
is caused by the strong FM couplings between neighboring Os and Ni atoms
($J_1$) as well as between neighboring Os and Eu atoms ($J_4$).
Compared with $J_1$ and $J_4$, the sizes of $J_3$ and \emph{J$_{5}$} are negligibly small.
The smallness of $J_3$ could be attributed to the much localized Ni 3$d$ orbitals and
also the much larger separation of the neighboring Ni atoms (see Table I).
The smallness of $J_5$ could result from the much weaker Ni 3$d$-Eu 4$\emph{f}$ hybridization
due to the much localized Ni 3$d$ as well as Eu 4$\emph{f}$ orbitals.
Interestingly, Table I indicates that \emph{J$_{3}$} and \emph{J$_{5}$} become positive in the (111) BaEuNiOsO$_6$ monolayer.
Table I also shows that in the monolayer, \emph{J$_{2}$} is enhanced by a factor of 4 and thus becomes dominant
while $J_1$ and $J_4$ get reduced by a factor of 2 and 3, respectively.
All these suggest that ferromagnetism in the monolayer is caused by strong FM couplings
between neighboring Os atoms ($J_2$) as well as between neighboring Os and Ni atoms
($J_1$) and between neighboring Os and Eu atoms ($J_4$).

In both BaEuNiOsO$_6$ structures, due to the presence of magnetic Eu atoms which are
staggered between the Ni and Os atoms, the number of magnetic coupling channels increases.
Compared with Ba$_{2}$NiOsO$_6$, the additional
coupling channels from Eu 4$\emph{f}$ to Os 5$d$ in the BaEuNiOsO$_6$
structures not only are of FM type (see Table I), thus enhancing the overall FM
exchange coupling strength, but also are the principal magnetic
couplings. This results in a substantial increase of $T_c$ and thus
the estimated $T_c$ value is as high as 325 K for bulk BaEuNiOsO$_6$ and 210 K
for its monolayer, being much larger than that of 150 K and 69 K in bulk Ba$_{2}$NiOsO$_6$
and its (111) monolayer~\cite{LuGuo19,Feng16}, respectively.
Nonetheless, it should be noted that the mean-field estimation of the magnetic transition temperature
works rather well only for the systems where the magnetic atoms are coupled to a large number
of neighboring magnetic atoms. This is often not the case for real magnetic materials,
and thus the transition temperature is generally overestimated by the mean-field approach (MFA),
especially for low-dimensional materials. We do notice that in an earlier study on bulk EuTiO$_3$ [54],
the GGA+U calculation plus the MFA correctly predicted bulk EuTiO$_3$ to
be a G-type antiferromagnetic insulator with $T_N$ = 6 K~\cite{lu15}.
%For 2D materials such as ferromagnetic monolayers (ML), because of their reduced magnetic atom coordination numbers,
%the MFA TC values are higher than the corresponding experimental values.
%On the other hand, in the ferromagnetic CrI$_3$ ML, the MFA $T_C$ is 109 K,
%being about twice as large as the experimental one of 45 K~\cite{lu}.
Therefore, we believe that the estimated $T_C$ value for bulk BaEuNiOsO$_6$ may be slightly too high,
while that of  its (111) ML superlattice would be significantly too high.

We notice that although magnetic topological insulating phases have been
proposed in many materials, so far there have been few experimental
reports on the observation of such topological phases\cite{xue,Weng15}.
Weak magnetic coupling and hence low $T_{c}$ values have been considered
as a key factor that hinders the experimental observation of these
topological phases.~\cite{Chandra17,Lu19} Therefore,
some schemes have been proposed to increase $T_{c}$.
One proposed route is to exploit the proximity
effect coupling the surface states of a nonmagnetic topological insulator
(TI) directly with a high $T_{c}$ magnetic insulator~\cite{Yan2020, Shi2017}.
An alternative route is to optimize the transition metals in the considered systems.
For example, it was proposed in Refs. ~\cite{Chandra17,Lu19}
to replace some B-site 3$d$ transition metal atoms in the (111) ABO$_{3}$ perovskite bilayers
with 4$d$ and 5$d$ atoms which
simultaneously have more extended $d$ orbitals and stronger SOC.
Here we offer another way to enhance
the $T_{c}$ by substitution of A-site atoms with 4$f$
magnetic atoms in transition metal perovskite oxides.

\subsection{Electronic band structure}

\begin{figure}
\includegraphics[width=7.5cm]{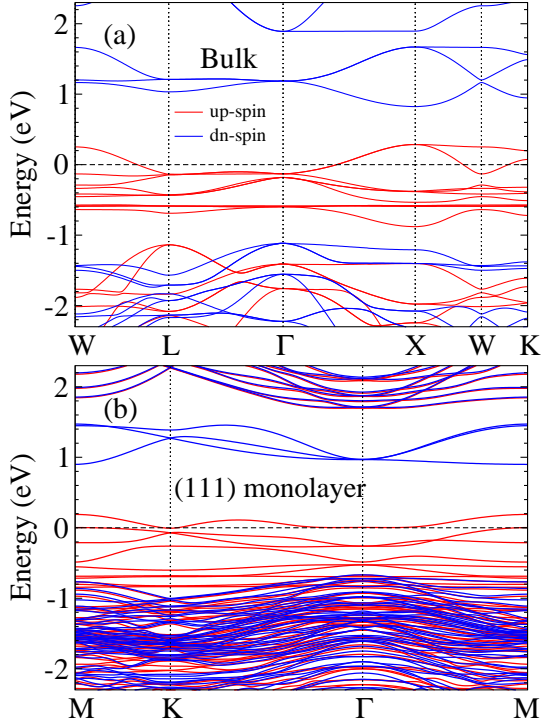}
\caption{Scalar-relativistic band structures of (a) bulk
BaEuNiOsO$_6$ and (b) the (111) (BaEuNiOsO$_{6}$)/(BaTiO$_{3}$)$_{10}$
superlattice obtained with the GGA+U method. Zero refers to the Fermi level.}
\end{figure}

The calculated scalar-relativistic band structures, atom- and
orbital-decomposed densities of states (DOSs) for both structures
are plotted in Figs. 5 and 6. These results are calculated by the GGA + U method with $U_{Os}$ = 2.0 eV, $U_{Ni}$ = 5.0 eV, and
$U_{Eu}$ = 4.0 eV.
In the absence of the SOC, both
BaEuNiOsO$_{6}$ structures are half-metallic. The energy
bands near the Fermi level are purely up-spin and made up of
the strongly hybridized Os 5$d$ $t_{2g}$, O $p$ and Eu 4$f$ orbitals.
The half-metallicity is consistent with the integer magnetic
moment of 11 $\mu_B$ per BaEuNiOsO$_{6}$ formula. The atom- and
orbital-decomposed DOSs of both structures show that Os 5$d$ $t_{2g}$ orbitals
overlap strongly with Eu 4$f$ orbitals and extend to the neighboring Ni
3$d$ orbitals through their hybridization with O $p$ orbitals.
This explains the large $J_{1}$ (Ni-Os) and $J_{4}$ (Eu-Os) values,
which give rise to the higher $T_c$, as mentioned above.

Generally speaking, the SOC has significant effect on the
electronic and magnetic properties of these 4$f$ and 5$d$ metal
perovskite materials. To find the magnetization orientation of
the FM state, we also perform the
total energy calculations with the SOC for magnetization
along the (001), (110), and (111) directions in bulk BaEuNiOsO$_6$,
and also for (001) and (100) directions in the (111) BaEuNiOsO$_6$ ML.
The easy magnetization axis is found along the (111) direction in bulk BaEuNiOsO$_6$.
Interestingly, the easy axis remains unchanged
in the (111) BaEuNiOsO$_6$ monolayer (along the $c$ axis, see Fig. 2), i.e.,
the (111) BaEuNiOsO$_6$ monolayer has a perpendicular anisotropy.
This may be useful for realizing high $T_c$ QAH phase in the
material.

\begin{figure}
\includegraphics[width=7cm]{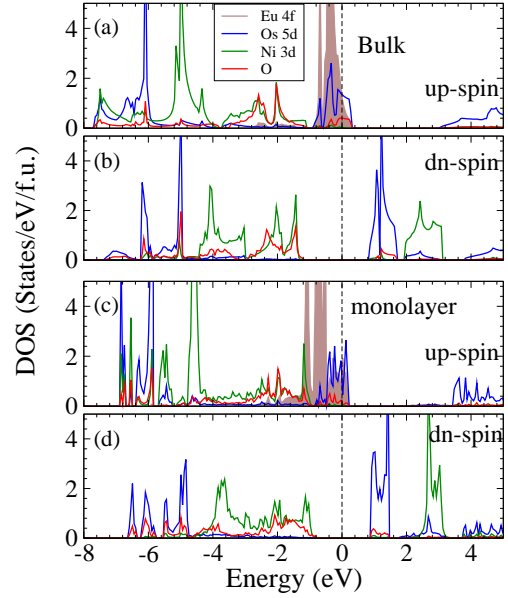}
\caption{Atom- and orbital-decomposed densities
of states (DOS) of (a, b) bulk BaEuNiOsO$_6$ and (c, d) the (111)
(BaEuNiOsO$_{6}$)/(BaTiO$_{3}$)$_{10}$ superlattice obtained with the GGA+U method.
Zero refers to the Fermi level.}
\end{figure}

The fully relativistic FM band structures
%, which is actually scalar-relativistic including SOC,
of both systems with the magnetization along the easy axis
are shown in Fig. 7. Interestingly, Fig. 7(a) indicates that
linear band crossings appear at the Fermi level
in bulk BaEuNiOsO$_6$ when the SOC is turned on. This makes BaEuNiOsO$_6$
a candidate for Weyl semimetal. On the other hand, when the SOC is switched on,
its (111) monolayer becomes an insulator with a small band gap of 93 meV,
thus making it a candidate for a FM topological insulator (or Chern insulator).

The calculated spin and orbital magnetic moments of both systems
in the FM state are listed in Table II, together with
Eu 4$f$-, Ni 3$f$-and Os 5$d$-shell configurations.
Due to the strong SOC effect, the calculated spin
magnetic moments fall short of the values expected from Os$^{6+}$
5$d^{2}$($\emph{t}_{2g}^{2}$ $\emph{e}_{g}^{0}$; $S=1$), Eu$^{2+}$
4$f^{7}$($\emph{f}^{7}$; $S=7/2$), and Ni$^{2+}$ 3$d^8$
($\emph{t}_{2g}^{6}$ $\emph{e}_{g}^{2}$; $S=1$) ions. In particular,
the calculated spin magnetic moment of an Os atom is 1.499
$\mu$$_{B}$ in bulk BaEuNiOsO$_6$ and 1.463 $\mu$$_{B}$ in its (111) ML.
Interestingly, all Ni, Os, and Eu
atoms have significant orbital magnetic moments, being 0.213
$\mu$$_{B}$, -0.367 and -0.473 $\mu$$_{B}$ in bulk BaEuNiOsO$_6$,
and 0.209 $\mu$$_{B}$, -0.345 $\mu$$_{B}$, -0.566 $\mu$$_{B}$ in the
(111) ML. Hund's second rule states that the spin and
orbital moments would be antiparallel if the $d$ or $f$ shell is
less than half-filled, and otherwise they would be parallel. In
consistence with Hund's second rule, the Ni orbital moment is
parallel to the Ni spin moment while the Os and Eu orbital moments
are antiparallel to their spin moments.

\begin{figure}
\includegraphics[width=7.6cm]{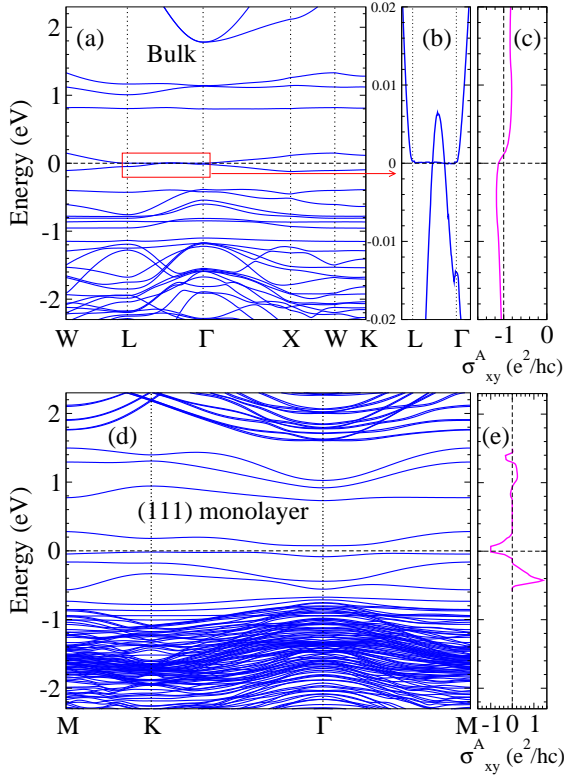}
\caption{(a,b,d) Relativistic
band structures and (c,d) anomalous Hall conductivity
($\sigma^A_{xy}$) of bulk BaEuNiOsO$_6$ (upper panels) and its
(111) (BaEuNiOsO$_{6}$)/(BaTiO$_{3}$)$_{10}$ superlattice (lower panels) obtained with the GGA+U+SOC method.
Zero refers to the Fermi level.}
\end{figure}

%\begin{figure}
%\includegraphics[width=7.5cm]{BaEuNiOsOFig8.eps}
%\caption{Charge density distributions of the upper valence
%bands (a, c) [energy window from -0.4 to $E_F$ (Fermi level)] and
%lower conduction bands (b, d) [energy range from
%$E_F$ to  $+0.4$ eV], without (upper panels) and with the SOC (lower panels) in
%the (111) (BaEuNiOsO$_{6}$)/(BaTiO$_{3}$)$_{10}$ superlattice.}
%\end{figure}

%In addition, the SOC would also greatly affect the orbital characters of
%the energy bands near the Fermi level. Take the (111) BaEuNiOsO$_6$ monolayer as
%an example (see Fig. 8). When the SOC is tuned on,
%we observe that some electrons in
%the upper valence bands in the BaO$_3$ layer
%[see Fig. 8(a)] are transferred to the adjacent Ti 3$d$ orbitals above [see Fig. 8(c)].
%Consequently, the Os 5$d$ $t_{2g}$ form covalent bonds with the neighboring O $p$
%via Ti 3$d$ orbitals [see Fig. 8(c)], and the contributions of the BaO$_3$ layer
%and the Os 5$d$ $t_{2g}$ to the lower conduction bands [see Fig. 8(b)] diminish [see Fig. 8(d)].
%In the meantime, there is also some charge transfer from the BaO$_3$ layer
%to the neighboring Ni 3$d$ orbitals below [see Fig. 8(c)]. This results in
%a strong overlap of the Os 5$d$ $t_{2g}$ with the neighboring O $p$ and Eu 4$f$ orbitals.
%Such strong covalent bonding among Os 5$d$ $t_{2g}$, Ti 3$d$, O 2$p$ and Ni 3$d$ orbitals along the
%$c$-axis may be the main reason that the magnetization prefers the out-of-plane direction in the
%(111) BaEuNiOsO$_6$ monolayer.

\begin{table}%[h]
\caption{Position coordinates (in units of $\frac{2\pi}{a_0}$) and chiralities
of the Weyl points in bulk BaEuNiOsO$_6$. Also listed is the momentum separation
$q^c$ (also in units of $\frac{2\pi}{a_0}$) %(modulo  $\frac{2\pi}{c}$)
along the [111] direction between the two Weyl points in each pair.
Here $a_0 = 8.0060$ \AA$ $ is the lattice constant of cubic BaEuNiOsO$_6$.}
\begin{ruledtabular}
\begin{tabular}{cccc}
pair &  positions                  & chirality & $q^c$ \\
  \hline
1 & (0.0740,    0.0740,   0.0740)  & +1 & -0.2563  \\
  & (-0.0740,   -0.0740,  -0.0740) & -1 &         \\
2 & (0.2349,    0.2349,    0.2349) & -1 & -0.9183  \\
  & (0.7651,    0.7651,    0.7651) & +1 &         \\
\end{tabular}
\end{ruledtabular}
\end{table}

\begin{figure}
\includegraphics[width=6.0cm]{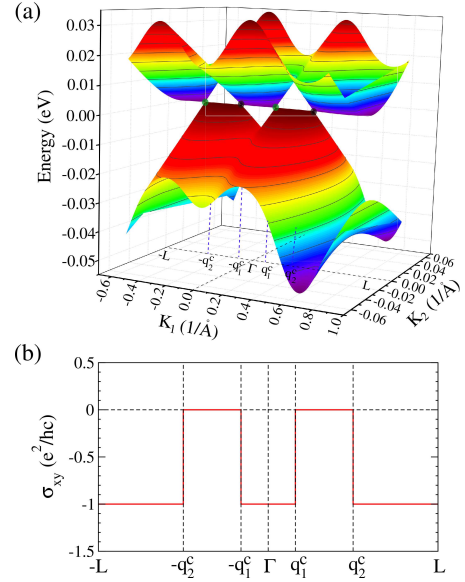}
\caption{
(a) Energy bands near the Weyl points on the $k$-plane containing the
$-L$-$\Gamma$-$L$ line
in the Brillouin zone (BZ) in bulk ferromagnetic BaEuNiOsO$_{6}$ with [111] magnetization.
(b) AHC ($\sigma_{xy}$) on the 2D $k$-plane perpendicular to [111] in the BZ as a function
of $k_{111}$ along the $-L$-$\Gamma$-$L$ line. See Table IV for the values
of critical $k$-points $q^c_1$ and $q^c_2$.
}
\end{figure}

\begin{figure}
\includegraphics[width=6.0cm]{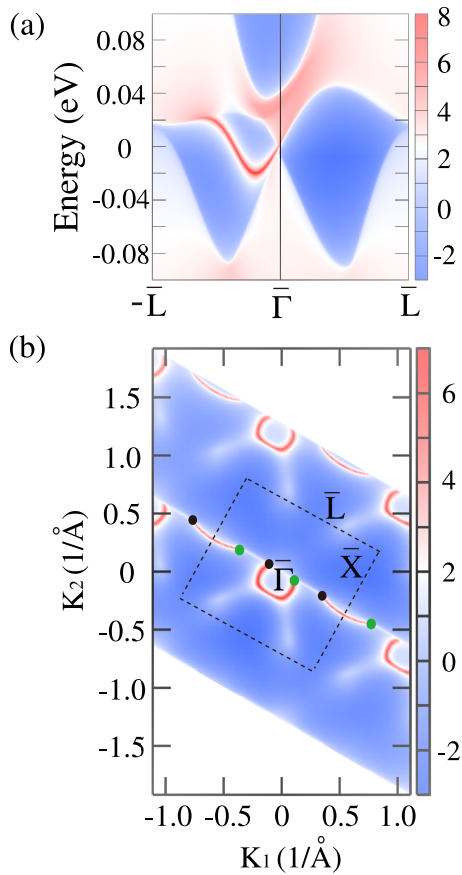}
\caption{The (001) surface states of bulk BaEuNiOsO$_6$.
(a) Energy dispersion of the surface states along high symmetry
line $-\overline{L}$-$\overline{\Gamma}$-$\overline{L}$
in the surface Brillouin zone (BZ) [see Fig. 1(c)].
(b) Surface state spectral function with the energy lying at Weyl points.
Here Weyl points with positive and negative chiralities are
represented by the black and green solid circles, respectively.
The dashed square indicates the surface BZ [see Fig. 1(c)].}
\end{figure}

\subsection{Magnetic Weyl semimetal phase}

As mentioned above in Sec. III (B), linear band crossings appear at
the Fermi level in bulk BaEuNiOsO$_6$
when the SOC is turned on [Fig. 7(a)], suggesting that bulk BaEuNiOsO$_6$
is a candidate for the magnetic WSM. One characteristic of a WSM
is the emergence of pairs of Weyl points with the two Weyl points in
each pair having opposite chiralities (chiral charges).
Therefore, we first determine the locations and chiralities
of the Weyl points in bulk BaEuNiOsO$_6$.
The position coordinates and chiralities of the Weyl points found in
bulk BaEuNiOsO$_6$ are listed in Table III, and their locations
and associated Weyl cones are displayed in Fig. 8. In contrast to many previously
predicted WSMs~\cite{ArmitageNP2018}, there are only two pairs
of Weyl points locating at $\pm$(0.0740,
0.0740, 0.0740) and $\pm$(0.2349, 0.2349, 0.2349), respectively.
Furthermore, they are all located on high symmetry line -L-$\Gamma$-L in the BZ.
Note that for the [111] magnetization in bulk BaEuNiOsO$_6$, there is a three-fold rotation symmetry
($C_{3v}$) along this -L-$\Gamma$-L  symmetry line.
Consequently, according to the two-band {\bf k}$\cdot${\bf p} theory analysis~\cite{FangC2012},
the four Weyl points found on this symmetry line are protected by the symmetry of
crystallographic point group $C_{3v}$. This is the same as
in the case of cubic spinel HgCr$_{2}$Se$_{4}$ with the magnetization along the [111] direction~\cite{FangC2012}.
However, HgCr$_{2}$Se$_{4}$ is not an ideal Weyl semimetal because
its Weyl points are located at $\sim$9 meV above the Fermi level
and also there are several nontopological bands at this energy level~\cite{Xug2011}.

Since Berry curvatures in the vicinity of the Weyl points are large~\cite{ArmitageNP2018},
one would expect a strong AHE in a magnetic WSM with the Fermi level
close to the Weyl points. Therefore, we compute the intrinsic AHC in bulk BaEuNiOsO$_6$
as a function of the Fermi level, and the result is displayed in Fig. 7(c).
Indeed, the calculated AHC at the Fermi level is large, being
-543 S/cm = -1.1225 $e^{2}/ha_0$ [see Fig. 7(c)] where $e^{2}/ha_0$ is the quantum of the Hall conductance
(here 1 $e^{2}/ha_0$ = 484 S/cm) and $a_0$ is the lattice constant of the cubic BaEuNiOsO$_6$.
In retrospect, this is not surprising since bulk BaEuNiOsO$_6$ is an ideal Weyl semimetal
with just two pairs of Weyl points and no other band crossing the Fermi level. It was shown
in Ref. \cite{Yang2011} that for an ideal magnetic Weyl semimetal,
the AHC would be given by
\begin{equation}
\sigma_{ab}^A = \epsilon_{abc}\frac{e^2}{2\pi h}\sum_i q^c_i,
\end{equation}
where $q^c_i$ is the [111] component of the momentum separation vector
of Weyl point pair $i$, and $i$ is summed over all Weyl point pairs.
Using the calculated  $q^c_i$ values listed in Table III,
we get $\sigma_{ab}^A = -1.1746 \frac{e^2}{ha_0} = -569$ S/cm,
in good agreement with the {\it ab initio} calculation.
Interestingly, Fig. 5(a) indicates that
bulk BaEuNiOsO$_6$ is half-metallic with the conduction bands being purely spin-up.
Therefore, we expect that the large anomalous Hall current in BaEuNiOsO$_6$
would be fully spin-polarized.

Another characteristic feature of a WSM is the occurrence of
the topological surface Fermi arcs. We thus calculate the band diagram
of the (001) surface of bulk BaEuNiOsO$_6$, as displayed along high-symmetry
line $-\overline{L}$-$\overline{\Gamma}$-$\overline{L}$ in Fig. 9(a).
One surface band connecting the bulk conduction and valence bands
along the $-\overline{L}$-$\overline{\Gamma}$ line can be clearly seen [see Fig. 9(a)].
We also display the spectral function of the surface states at
the energy of the Weyl points over the surface BZ in Fig. 9(b).
Perfect surface Fermi arc states can clearly be seen and the surface Fermi arcs begin and
end at the Weyl points with opposite chirality (labeled by the
green and black circles) in Fig. 9(b). Importantly, the surface Fermi
arc that connects the Weyl points of pair 2 is long, extending
about 68 \% of the [111] reciprocal lattice vector (see Table III).
It is worthwhile emphasizing that the Weyl points are located right at the Fermi level.
Therefore, the WSM states in bulk BaEuNiOsO$_6$ could be easily
detected by such surface techniques such as ARPES and scanning tunneling microscopy (STM).
More importantly, many exotic quantum phenomena predicted for WSMs~\cite{ArmitageNP2018} should
be observed in FM BaEuNiOsO$_6$ at high temperature because it is an ideal WSM.
We also display the calculated band diagram of the (111) surface in Fig. S2(a) in
the SM~\cite{SM} and also the spectral function of the surface states at the Fermi level
in Fig. S2(b) in the SM. Figure S2(a) shows that one surface band crossing the Fermi level
and connecting the bulk conduction and valence bands along both the $\overline{M}$-$\overline{\Gamma}$
and $\overline{\Gamma}$-$\overline{M}$ lines.
However, there are only closed circular surface states [Fig. S2(b)].
There is no surface Fermi arc on this surface because all the Weyl points
which are on the -L-$\Gamma$-L line, now are underneath the surface BZ center (the $\overline{\Gamma}$ point)
[see Fig. 1(d)].

\begin{figure}
\includegraphics[width=7cm]{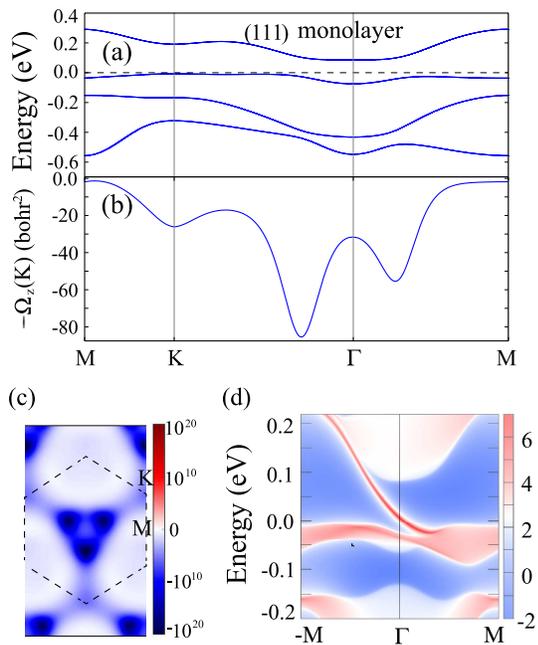}
\caption{(a) Relativistic band structure near the Fermi level,
(b) Berry curvature along the symmetry lines and
(c) distribution of the Berry curvature on the $k_{z}$ = 0 plane in the BZ
of the (111) BaEuNiOsO$_{6}$ ML in the (BaEuNiOsO$_{6}$)/(BaTiO$_{3}$)$_{10}$ superlattice.
(d) The edge band diagram of the ML with the edge along the (010) direction.
}
\end{figure}

\subsection{Chern insulator phase}

As mentioned above, the (111) BaEuNiOsO$_6$ ML is
half-metallic in the absence of the SOC [Fig. 5(b)].
However, it becomes a semiconductor when the SOC is included.
Thus, we could expect that the SOC-opened band gap in the ML
would be topologically nontrivial. To examine
the topological nature of the band gap, we calculate the AHC ($\sigma^A_{xy}$)
[see Eq. (1)] of this monolayer. For a Chern insulator,
$\sigma^A_{xy} =  n e^2/hc$ where $c$ is the lattice constant
along the $c$-axis normal to the plane of longitudinal and Hall
currents and $n$ is an integer known as the Chern
number~\cite{Hal87,Zhou2016}. For a normal FM insulator, on the other hand,
$\sigma^A_{xy} = 0$. The calculated AHC of the (111) BaEuNiOsO$_{6}$
ML is displayed as a function of the Fermi level in Fig. 7(e).
Interestingly, when the Fermi level falls within the band gap,
$\sigma^A_{xy} =  -1 e^2/hc$ [Fig. 7(e)]. This shows that it is
a Chern insulator with Chern number $n = -1$.
Equation (1) shows that the nonzero Chern number is due to the nonzero
Berry curvature of the occupied bands caused by the simultaneous
occurrence of the magnetization and SOC~\cite{XiaoD10}.
To locate the hot spots of the Berry curvature in the BZ,
we display the Berry curvature along the symmetry
lines in Fig. 10(b) and also on the $k_z =0$ plane in Fig. 10(c).
Figure 10(b) shows that the Berry curvature has a prominent negative peak along the
K-$\Gamma$ line near the $\Gamma$ point. In fact, because the BZ has a three-fold
rotation axis at the $\Gamma$ point, there are three such peaks in
the vicinity of the $\Gamma$ point [see Fig. 10(c)].

According to the bulk-edge correspondence theorem, the monolayer
should have one chiral gapless edge mode since its Chern number $n = -1$.
To verify this finding, we calculate the edge energy bands along the (010) direction,
as spectral functions plotted in Fig. 10(d). There is clearly one metallic edge band
emerging within the band gap, being in accordance with the calculated Chern number
$n = -1$. Furthermore, this topological band gap is opened in
the purely up-spin bands [see Fig. 5(b)]. Thus, it is a single-spin topological
semiconducting phase and the edge current would be fully spin-polarized~\cite{Liang}.
In retrospect, that the (111) BaEuNiOsO$_{6}$ ML superlattice
is a Chern insulator should not come as a surprise because
it is the (111) quantum-well structure of bulk BaEuNiOsO$_{6}$.
Like cubic spinel HgCr$_2$Se$_4$~\cite{Xug2011}, bulk FM BaEuNiOsO$_{6}$
is a Chern semimetal. Thus, as a unique physical result characterizing
a Chern semimetal, the (111) BaEuNiOsO$_{6}$ ML, being
a quantum-well structure of bulk BaEuNiOsO$_{6}$, should be a Chern insulator.~\cite{Xug2011}

To ensure that this topological property found in the (111) BaEuNiOsO$_{6}$ ML
is robust against the thickness variation of the
insulating BaTiO$_3$ layer, we further perform the
calculations for the (111) (BaEuNiOsO$_{6}$)/(BaTiO$_3$)$_{7}$,
(BaEuNiOsO$_{6}$)/(BaTiO$_3$)$_{13}$ and
(BaEuNiOsO$_{6}$)/(BaTiO$_3$)$_{16}$ superlattices.
We find that all the three superlattices have a band structure almost identical
to that of the (111) (BaEuNiOsO$_{6}$)/(BaTiO$_3$)$_{10}$ with a
slightly different band gap (see Fig. S3 in the SM~\cite{SM}).
Considering the (111) BaEuNiOsO$_{6}$ ML could be grown
on other insulating oxide substrates such as SrTiO$_{3}$, NaTaO$_{3}$ and KNbO$_{3}$,
we also carry out the calculations for the (111)
(BaEuNiOsO$_{6}$)/(BaTiO$_3$)$_{10}$ superlattice with the
in-plane lattice strains of -2 \%, -1 \% and 0.71 \%,
which correspond, respectively, to the possible lattice mismatch strains
when grown on SrTiO$_{3}$, NaTaO$_{3}$ and KNbO$_{3}$
with the theoretically determined cubic lattice constants being
3.946 \AA, 3.985 \AA$ $ and 4.057 \AA. We again find that all the strained superlattices
have a nearly identical band structure with a slightly changed band
gap compared to that of the unstrained (BaEuNiOsO$_{6}$)/(BaTiO$_3$)$_{10}$
(see Fig. S4 in the SM~\cite{SM}).

\section{Conclusions}

In conclusion, based on systematic first-principles DFT calculations,
we find that the (111) ML (BaEuNiOsO$_{6}$)/(BaTiO$_3$)$_{10}$
superlattice is a FM Chern insulator with high $T_C$ of about 210 K
and also a large band gap of $\sim$90 meV.
Guided by the strategy of "inverse engineering", we uncover
the corresponding bulk BaEuNiOsO$_{6}$ to be an ideal
FM WSM with above room temperature $T_C$ of around 325 K and two pairs of Weyl points at
the Fermi level only. Furthermore, one pair of Weyl points has a large momentum separation,
thus giving rise to not only a long (001) surface Fermi arc but also large AHC of 543 S/cm.
Analysis of the calculated band structure and also exchange coupling parameters
reveal that the strong ferromagnetism in these double perovskite BaEuNiOsO$_{6}$
structures is driven by not only the rare nearest-neighbor Ni-Os FM coupling but also Eu-Os FM coupling.
Therefore, double perovskite BaEuNiOsO$_{6}$ will offer
a superior material platform for exploring exotic Weyl fermion physics
and its (111) ML superlattices will provide a high temperature quantum phenomena such as quantum anomalous Hall effect.
This work is thus expected to stimulate further experimental and theoretical investigations
on these interesting materials.

\begin{acknowledgments}
H.-S.L acknowledges the support from the National Natural
Science Foundation of China under Grant No. 11704046. G.-Y. G.
thanks the support from the Ministry of Science and Technology and
the National Center for Theoretical Sciences in Taiwan.
\end{acknowledgments}


\begin{thebibliography}{99}

\bibitem{Weyl29} H. Weyl, Z. Phys. \textbf{56}, 330 (1929).

\bibitem{Wanxg2011} X. G. Wan, A. M. Turner, A. Vishwanath, and S. Y. Savrasov, Phys.
Rev. B \textbf{83}, 205101 (2011).

\bibitem{Shiw2018} W. Shi, L. Muechler, K. Manna, Y. Zhang, K. Koepernik, R. Car, J.
v. d. Brink, C. Felser, and Y. Sun, Phys. Rev. B \textbf{97},
060406(R) (2018).

\bibitem{Volovik} G. E. Volovik, The Universe in A Helium Droplet (Clarendon Press,
Oxford, 2003).

\bibitem{YanB2017} B. Yan and C. Felser, Annu. Rev. Condens. Matter Phys. \textbf{8}, 337 (2017).

\bibitem{ArmitageNP2018} N. P. Armitage, E. J.Mele, and A. Vishwanath, Rev. Mod. Phys. \textbf{90},
15001 (2018).

\bibitem{Hasanmz2010} M. Z. Hasan and C. L. Kane, Rev. Mod. Phys. \textbf{82}, 3045
(2010).

\bibitem{DingH2015} B. Q. Lv, H. M. Weng, B. B. Fu, X. P.Wang, H. Miao, J. Ma, P.
Richard, X. C. Huang, L. X. Zhao, G. F. Chen, Z. Fang, X. Dai, T.
Qian, and H. Ding, Phys. Rev. X \textbf{5}, 031013 (2015).

\bibitem{YangLX2015} L. X. Yang, Z. K. Liu, Y. Sun, H. Peng, H. F. Yang, T. Zhang, B.
Zhou, Y. Zhang, Y. F. Guo, M. Rahn, D. Prabhakaran, Z. Hussain, S.
K. Mo, C. Felser, B. Yan, and Y. L.Chen, Nat. Phys. \textbf{11},
728 (2015).

\bibitem{LiuZK2016} Z. K. Liu, L. X. Yang, Y. Sun, T. Zhang, H. Peng, H. F. Yang, C.
Chen, Y. Zhang, Y. F. Guo, D. Prabhakaran, M. Schmidt, Z. Hussain,
S.-K. Mo, C. Felser, B.Yan, and Y. L. Chen, Nat.Mater.
\textbf{15}, 27 (2016).

\bibitem{Hasanmz2015} S.-Y. Xu, I. Belopolski, N. Alidoust, M. Neupane, G. Bian, C.
Zhang, R. Sankar, G. Chang, Y. Zhujun, C.-C. Lee, H. Shin- Ming,
H. Zheng, J. Ma, D. S. Sanchez, B. Wang, A. Bansil, F. Chou, P. P.
Shibayev, H. Lin, S. Jia, and M. Z. Hasan, Science \textbf{349},
613 (2015).

\bibitem{Shim2016} N. Xu, H. M. Weng, B. Q. Lv, C. E. Matt, J. Park, F. Bisti, V. N.
Strocov, D. Gawryluk, E. Pomjakushina, K. Conder, N. C. Plumb, M.
Radovic, G. Autes, O. V. Yazyev, Z. Fang, X. Dai, T. Qian, J.
Mesot, H. Ding, and M. Shi, Nat. Commun. \textbf{7}, 11006 (2016).

\bibitem{Inoueh2016} H. Inoue, A. Gyenis, Z. Wang, J. Li, S. W. Oh, S. Jiang, N. Ni, B.
A. Bernevig, and A. Yazdani, Science 351, 1184 (2016).

\bibitem{HuangChen2015} X. Huang, L. Zhao, Y. Long, P. Wang, D. Chen, Z. Yang, H. Liang,
M. Xue, H. Weng, Z. Fang, X. Dai, and G. Chen, Phys. Rev. X
\textbf{5}, 031023 (2015).

\bibitem{ZhangJia2016} C.-L. Zhang, S.-Y. Xu, I. Belopolski, Z. Yuan, Z. Lin, B. Tong, G.
Bian, N. Alidoust, C.-C. Lee, S.-M. Huang, T.-R. Chang, G. Chang,
C.-H. Hsu, H.-T. Jeng, M. Neupane, D. S. Sanchez, H. Zheng, J.
Wang, H. Lin, C. Zhang, H.-Z. Lu, S.-Q. Shen, T. Neupert, M. Zahid
Hasan, and S. Jia, Nat. Commun. \textbf{7}, 10735 (2016).

\bibitem{LiuFelser2018} E. Liu, Y. Sun, N. Kumar, L. Muechler, A. Sun, L. Jiao, S. Yang,
D. Liu, A. Liang, Q. Xu, J. Kroder, V. S\"{u}$\beta$, H. Borrmann,
C. Shekhar, Z. Wang, C. Xi, W. Wang, W. Schnelle, S. Wirth, Y.
Chen, S. T. B. Goennenwein, and C. Felser, Nat. Phys. \textbf{14},
1125 (2018).

\bibitem{Ahn2020} J. Ahn, G.-Y. Guo and N. Nagaosa, Phys. Rev. X \textbf{10}, 041041 (2020).

\bibitem{Xug2011} G. Xu, H. Weng, Z. Wang, X. Dai, and Z. Fang, Phys. Rev. Lett.
\textbf{107}, 186806 (2011)

\bibitem{TonyLow} W. Jiang, H. Huang, F. Liu, J. Wang, and
T. Low, Phys. Rev. B \textbf{101}, 121113(R) (2020)

\bibitem{Hirschberger2016} M. Hirschberger, S. Kushwaha, Z. Wang, Q. Gibson, S. Liang, C. A Belvin, B A Bernevig, R J Cava, N
P Ong, Nat. Mater. \textbf{15}, 1161 (2016).

\bibitem{Shekhar2018} C. Shekhara, N. Kumar, V. Grinenko, S. Singh, R. Sarkar, H. Luetkens, S. Wu, Y. Zhang,
A. C. Komarek, E. Kampert, Y. Skourski, J. Wosnitza, W. Schnelle,
A. McCollam, U. Zeitler, J. K¨¹bler, B. Yan, H.-H. Klauss, S. S.
P. Parkin, and C. Felser, Proc. Natl Acad. Sci. USA \textbf{115},
9140 (2018).


\bibitem{KublerJ2016} J. Kubler, and C. Felser, Europhys. Lett. \textbf{114}, 47005
(2016).

\bibitem{Wangz2016} Z. Wang, M. G. Vergniory, S. Kushwaha, Max Hirschberger, E. V.
Chulkov, A. Ernst, N. P. Ong, Robert J. Cava, and B. Andrei
Bernevig, Phys. Rev. Lett. \textbf{117}, 236401 (2016)

\bibitem{Changg2016} G. Chang, S. Y. Xu, H. Zheng, B. Singh, C. H.
Hsu, G. Bian, N. Alidoust, I. Belopolski, D. S. Sanchez, S. Zhang,
H. Lin, and M. Z. Hasan, Sci. Rep. \textbf{6}, 38839 (2016).

\bibitem{HasanMZ19} I. Belopolski, K. Manna, D. S. Sanchez, G. Chang, B. Ernst, J. Yin, S. S. Zhang, T. Cochran, N. Shumiya,
H.Zheng, B. Singh, G. Bian, D. Multer, M. Litskevich, X. Zhou, S. Huang, B. Wang, T. Chang, S. Xu, A. Bansil, C. Felser, H. Lin,
M. Z. Hasan, Science \textbf{365}, 1278 (2019).

%\bibitem{LiuFelser18} E. Liu, Y. Sun, N. Kumar, L. Muechler, A.
%Sun, L. Jiao, S. Yang, D. Liu, A. Liang, Q. Xu, J. Kroder, V. S¨¹,
%H. Borrmann, C. Shekhar, Z. Wang, C. Xi, W. Wang, W. Schnelle, S.
%Wirth, Y. Chen, S. T. B. Goennenwein and C. Felser, Nat. Phys.
%\textbf{14}, 1125¨C1131 (2018).

\bibitem{ChenYL19} D. F. Liu, A. J. Liang, E. K. Liu, Q. N. Xu, Y. W. Li, C. Chen, D. Pei, W. J. Shi, S. K. Mo, P. Dudin, T. Kim, C. Cacho, G. Li,
                   Y. Sun, L. X. Yang, Z. K. Liu, S. S. P. Parkin, C. Felser and Y. L. Chen, Science \textbf{365}, 1282 (2019).

\bibitem{Beidenkopf19} N. Morali, R. Batabyal, P. K. Nag, E. Liu, Q. Xu, Y. Sun, B. Yan, C. Felser, N. Avraham and H. Beidenkopf, Science \textbf{365}, 1286 (2019).

\bibitem{ChangG2018} G. Chang, B. Singh, S. Xu, G. Bian, S. Huang, C. Hsu, I. Belopolski, N. Alidoust, D. S. Sanchez, H. Zheng, H. Lu, X. Zhang, Y. Bian, T. Chang, Horng-Tay Jeng, A. Bansil,
H. Hsu, S. Jia, T. Neupert, H. Lin, M. Zahid Hasan, Phys. Rev. B
\textbf{97}, 041104(R) (2018).

\bibitem{Sanchez2020} D. S. Sanchez, G. Chang, I. Belopolski, H. Lu, J.-X. Yin, N. Alidoust, X. Xu, T. A. Cochran,
X. Zhang, Y. Bian, S. S. Zhang, Y.-Y. Liu, J. Ma, G. Bian, H. Lin, S.-Y. Xu, S. Jia and M. Z. Hasan,
Nature Commun. \textbf{11}, 3356 (2020).

\bibitem{Soh2019} J. R. Soh, F. de Juan, M. G. Vergniory, N. B. M. Schr\"{o}ter, M. C. Rahn, D. Y. Yan,
J. Jiang, M. Bristow, P. Reiss, J. N. Blandy, Y. F. Guo, Y. G. Shi, T. K. Kim, A. McCollam, S. H. Simon,
Y. Chen, A. I. Coldea, and A. T. Boothroyd, Phys. Rev. B \textbf{100}, 201102(R) (2019).

\bibitem{Kobayashi} K.-I. Kobayashi, T. Kimura, H. Sawada, K. Terakura and Y. Tokura,
Nature (London) \textbf{395}, 677 (1998).

\bibitem{Mannhart} J. Mannhart and D. G. Schlom, Science \textbf{327}, 1607 (2010).

\bibitem{Hwang12} H. Y. Hwang, Y. Iwasa, M. Kawasaki, B. Keimer, N. Nagaosa and Y. Tokura,
 Nature Mater. \textbf{11}, 103 (2012)

\bibitem{Xiao11} D. Xiao, W. Zhu, Y. Ran, N. Nagaosa, and S. Okamoto, Nat. Commun. \textbf{2}, 596 (2011).

\bibitem{Weng15} H. M. Weng, R. Yu, X. Hu, X. Dai, and Z. Fang, Adv. Phys. 64, 227 (2015).

\bibitem{Ran11} K.-Y. Yang, W. Zhu, D. Xiao, S. Okamoto, Z. Wang, and
Y. Ran, Phys. Rev. B \textbf{84}, 201104(R) (2011).

\bibitem{Dong15} Y. Weng, X. Huang, Y. Yao, and S. Dong, Phys. Rev. B \textbf{92}, 195114 (2015).

\bibitem{Dai15} Y. Wang, Z. Wang, Z. Fang, and X. Dai, Phys. Rev. B \textbf{91}, 125139 (2015).

\bibitem{Held17} L. Si, O. Janson, G. Li, Z. Zhong, Z. Liao, G. Koster, and K. Held, Phys. Rev. Lett. \textbf{119}, 026402 (2017).

\bibitem{Lu19} H.-S. Lu, and G.-Y. Guo, Phys. Rev. B \textbf{99}, 104405 (2019).

\bibitem{Chandra17} H. K. Chandra, and G.-Y. Guo, Phys. Rev. B \textbf{95}, 134448 (2017).

\bibitem{Cook14} A. M. Cook and A. Paramekanti, Phys. Rev. Lett. \textbf{113}, 077203 (2014).

\bibitem{Baidya16} S. Baidya, U. V. Waghmare, A. Paramekanti, and T. SahaDasgupta, Phys. Rev. B \textbf{94}, 155405 (2016).

%\bibitem{Haldane} F. D. M. Haldane, Phys. Rev. Lett. \textbf{61}, 2015 (1988).

\bibitem{ChenY2013} Y. Chen, D. L. Bergman, and A. A. Burkov, Phys. Rev. B \textbf{88}, 125110 (2013).

\bibitem{Yang2011} K.-Y. Yang, Y.-M. Lu and Y. Ran, Phys. Rev. B \textbf{84}, 075129 (2011).

%\bibitem{TKNN} D. J. Thouless, M. Kohmoto, M. P. Nightingale, and M. den Nijs, Phys. Rev. Lett. \textbf{49}, 405 (1982).

%\bibitem{Lau83} R. B. Laughlin, Phys. Rev. Lett. \textbf{50}, 1395 (1983).

%\bibitem{SunY2020} L. Muechler, E. Liu, J. Gayles, Q. Xu, C. Felser and Y. Sun, Phys. Rev. B \textbf{101}, 115106 (2020).

%\bibitem{Daix20} S. Nie, Y. Sun, F. B. Prinz, Z. Wang, H. Weng, Z. Fang and X. Dai, Phys. Rev. Lett. \textbf{124}, 076403 (2020).

\bibitem{LuGuo19} H.-S. Lu, and G.-Y Guo, Phys. Rev. B \textbf{100}, 054443 (2019).

\bibitem{Feng16} H. L. Feng, S. Calder, M. P. Ghimire, Y. H. Yuan, Y. Shirako, Y. Tsujimoto, Y. Matsushita,
Z. Hu, C.-Y. Kuo, L. H.  Tjeng, T.-W. Pi, Y.-L. Soo, J. He, M.
Tanaka, Y. Katsuya, M. Richter, and K. Yamaura, Phys. Rev. B
\textbf{94}, 235158 (2016).

\bibitem{Wang09} Y. K. Wang, P. H. Lee and G. Y. Guo, Phys. Rev. B \textbf{80}, 224418 (2009).

\bibitem{SM} See Supplemental Material at http://link.aps.org/supplemental/ for supplementary note 1,
Tables S1-S2 and Figs. S1-S4.

\bibitem{Perdew96} J. P. Perdew, K. Burke, and M. Ernzerhof, Phys. Rev. Lett. \textbf{77}, 3865 (1996).

\bibitem{PEB} P. E. Bl\"{o}chl, Phys. Rev. B \textbf{50}, 17953 (1994).

\bibitem{Kresse93} G. Kresse and J. Hafner, Phys. Rev. B \textbf{47}, 558
(1993); G. Kresse and J. Furthm\"{u}ller, Phys. Rev. B
\textbf{54}, 11169 (1996).

\bibitem{dudarev98} S. L. Dudarev, G. A. Botton, S. Y. Savrasov, C. J. Humphreys and A. P. Sutton, Phys. Rev. B 57, 1505 (1998).

\bibitem{lu15} H. S. Lu, T.-Y. Cai, S. Ju, and C. D. Gong, Phys. Rev. Applied \textbf{3}, 034011 (2015).

\bibitem{XiaoD10} D. Xiao, M.-C. Chang, and Q. Niu, Rev. Mod. Phys. \textbf{82}, 1959 (2010).

\bibitem{WangX06} X. Wang, J. R. Yates, I. Souza, and D. Vanderbilt, Phys. Rev.
B \textbf{74}, 195118 (2006).

\bibitem{LopezMG} M. G. Lopez, D. Vanderbilt, T. Thonhauser, and I. Souza, Phys. Rev. B \textbf{85}, 014435 (2012).

\bibitem{MarzariN} N. Marzari, A. A. Mostofi, J. R. Yates, I. Souza, and D. Vanderbilt, Rev. Mod. Phys. \textbf{84}, 1419 (2012).

\bibitem{Rubio84} M. P. Lopez Sancho, J. M. Lopez Sancho, and J. Rubio, J. Phys. F \textbf{14}, 1205 (1984).

\bibitem{Rubio85} M. P. Lopez Sancho, J. M. Lopez Sancho, J. M. L. Sancho, and J. Rubio, J. Phys. F \textbf{15}, 851 (1985).

\bibitem{Soluyanov18} Q. S. Wu, S. N. Zhang, H.-F. Song, M. Troyer, and A. A. Soluyanov, Comput. Phys. Commun. \textbf{224}, 405 (2018).

\bibitem{xue} C.-Z. Chang, J. Zhang, X. Feng, J. Shen, Z. Zhang, M. Guo, K. Li, Y. Ou, P. Wei, L.-L. Wang,
              Z.-Q. Ji, Y. Feng, S. Ji, X. Chen, J. Jia, X. Dai, Z. Fang, S.-C. Zhang, K. He, Y. Wang,
              L. Lu, X.-C. Ma, Q.-K. Xue, Science \textbf{340}, 167 (2013).

\bibitem{Yan2020} H. X. Fu, C. X. Liu, B. H. Yan, Sci. Adv. \textbf{6}, eaaz0948 (2020).

\bibitem{Shi2017} C. Tang, C.-Z. Chang, G. Zhao, Y. Liu, Z. Jiang, C.-X. Liu, M. R. McCartney, D. J. Smith, T. Chen,
J. S. Moodera, J. Shi, Sci. Adv. \textbf{3}, e1700307 (2017).

%\bibitem{Lu19} H. S. Lu, and G. Y. Guo, Phys. Rev. B \textbf{99}, 104405 (2019).

%\bibitem{Chandra17} H. K. Chandra, and G. Y. Guo, Phys. Rev. B \textbf{95}, 134448 (2017).

\bibitem{FangC2012} C. Fang, M. J. Gilbert, X. Dai, and B. A. Bernevig, Phys. Rev. Lett. \textbf{108}, 266802 (2012).

\bibitem{Hal87} B. I. Halperin, Jpn. J. Appl. Phys. \textbf{26}, 1913 (1987).

\bibitem{Zhou2016} J. Zhou, Q.-F. Liang, H. Weng, Y. B. Chen, S.-H. Yao, Y.-F. Chen, J. Dong and G.-Y. Guo,
Phys. Rev. Lett. \textbf{116}, 256601 (2016).

\bibitem{Liang} Q.-F. Liang, L.-H. Wu and X. Hu, New J. Phys. \textbf{15}, 063031 (2013).

%\bibitem{Hall79} E. H. Hall, Am. J. Math. \textbf{2}, 287 (1879).

%\bibitem{Hall81} E. H. Hall, Philos. Mag. \textbf{12}, 157 (1881).

%\bibitem{Klitzing} K. V. Klitzing, G. Dorda, and M. Pepper, Phys. Rev. Lett. \textbf{45}, 494 (1980).

%\bibitem{Liu08} C. X. Liu, X.-L. Qi, X. Dai, Z. Fang and S.-C. Zhang, Phys. Rev. Lett. \textbf{101}, 146802 (2008).

%\bibitem{ryu} R. Yu, W. Zhang, H. J. Zhang, S. C. Zhang, X. Dai, and Z. Fang, Science \textbf{329}, 61 (2010).

%\bibitem{Qia10} Z. H. Qiao, S. A. Yang, W. X. Feng, W.-K. Tse, J. Ding, Y. G. Yao, J. Wang and Q. Niu,
% Phys. Rev. B \textbf{82}, 161414(R) (2010).

%\bibitem{Che11} T.-W. Chen, Z.-R. Xiao, D.-W. Chiou and G. Y. Guo, Phys. Rev. B \textbf{84}, 165453 (2011).

%\bibitem{Ju2015} T.-Y. Cai, X. Li, F. Wang, S. Ju, J. Feng and C. D. Gong, Nano Lett. \textbf{15}, 6434-6439 (2015).

%\bibitem{Zhou16} J. Zhou, Q.-F. Liang, H. Weng, Y. B. Chen, S.-H. Yao, Y.-F. Chen, J. Dong and G. Y. Guo
%Phys. Rev. Lett. \textbf{116}, 256601(2016).

\end{thebibliography}
\end{document}